\documentclass[aps,twocolumn,prl,floatfix,superscriptaddress,showpacs,lengthcheck,hyperref,citeautoscript]{revtex4-1}
\usepackage{amsmath}
\usepackage{amssymb}
\usepackage{graphicx}
\usepackage{bm}
\usepackage{color}
\usepackage{times}
\usepackage{MnSymbol}
\usepackage{verbatim}

\newcommand{\bea}{\begin{eqnarray}}
\newcommand{\eea}{\end{eqnarray}}
\newcommand{\be}{\begin{equation}}
\newcommand{\ee}{\end{equation}}

\newcommand{\ket}[1]{| #1 \rangle}
\newcommand{\bra}[1]{\langle #1 |}

\begin{document}
\title{Photo-electrons unveil topological transitions in graphene-like systems}
\author{Lucila Peralta Gavensky}
\affiliation{Centro At{\'{o}}mico Bariloche and Instituto Balseiro,
Comisi\'on Nacional de Energ\'{\i}a At\'omica, 8400 Bariloche, Argentina}

\author{Gonzalo Usaj}
\affiliation{Centro At{\'{o}}mico Bariloche and Instituto Balseiro,
Comisi\'on Nacional de Energ\'{\i}a At\'omica, 8400 Bariloche, Argentina}
\affiliation{Consejo Nacional de Investigaciones Cient\'{\i}ficas y T\'ecnicas (CONICET), Argentina}

\author{C. A. Balseiro}
\affiliation{Centro At{\'{o}}mico Bariloche and Instituto Balseiro,
Comisi\'on Nacional de Energ\'{\i}a At\'omica, 8400 Bariloche, Argentina}
\affiliation{Consejo Nacional de Investigaciones Cient\'{\i}ficas y T\'ecnicas (CONICET), Argentina}

\begin{abstract}
The topological structure of the wavefunctions of particles in periodic potentials is characterized by the Berry curvature $\Omega_{kn}$ whose integral on the Brillouin zone is a topological invariant known as the Chern number. The bulk-boundary correspondence states that these numbers define the number of edge or surface topologically protected states. It is then of primary interest to find experimental techniques able to measure the Berry curvature. However, up to now, there are no spectroscopic experiments that proved to be capable to obtain information on $\Omega_{kn}$ to distinguish different topological structures of the $\it{bulk}$ wavefunctions of semiconducting materials. Based on experimental results of the dipolar matrix elements for graphene, here we show that ARPES experiments with the appropriate x-ray energies and polarization can unambiguously detect changes of the Chern numbers in dynamically driven graphene and graphene-like materials opening new routes towards the experimental study of topological properties of condensed matter systems. 
\end{abstract}
\maketitle

Topology plays a central role in defining the structure of the ground state of condensed matter systems, the nature of the excitations and their response to external probes~\cite{Hasan2010,Ando2013,Bernevig2013,Shen2013}. 
For particles in periodic potentials, like electrons in solids, cold atoms systems or photonic crystals, the topology of the Bloch wavefunctions determines the geometric or Berry phase acquired by the particle as it moves along a closed path in reciprocal space~\cite{Xiao2010}. Within a given energy band, these phases are characterized by the Berry curvature ($\Omega_{\bm{k}n}$) whose integral over the Brillouin zone (BZ) is a topological invariant, the Chern number.  

According to the bulk-boundary correspondence principle, the Chern numbers determine the unbalance in the number of chiral edge (or surface) states~\cite{Hasan2010}. Experimentally, it has been easier to study the effects of a non-trivial topology, \textit{i.e.} the emergence of such chiral edge states, rather than its origin: the structure of the Bloch wavefunctions across the whole BZ. In fact transport and spectroscopic experiments provide direct evidence on the existence of the edge states \cite{Koenig2007,Hsieh2008,Karch2011, Kristinsson2016}.  Extracting information on the Berry curvature and its integral in the BZ as a measure of topology in condensed matter systems has been more elusive.

Non-trivial topologies may be generated by external magnetic fields, spin-orbit coupling or by dynamically driving a system with external time dependent fields~\cite{Oka2009,Kitagawa2010,Lindner2011,Rudner2013,Perez-Piskunow2014,Usaj2014a}. The latter creates a new class of topological insulators known as Floquet Topological Insulators (FTI).

Since it is $\Omega_{\bm{k}n}$ what encodes all the information on topology and non-local effects---even in systems with trivial topology  $\Omega_{\bm{k}n}$ is associated with anomalous velocities~\cite{Chang1995,Sundaram1999,Xiao2010} and may lead to non-local conductances and unconventional (valley) Hall effects---, it is natural to look for ways of obtaining direct information about this quantity. Ultra-cold atoms in optical lattices offer a unique playground for the study of topological band structures \cite{Tarruell2012} and during the last years a number of experiments focused on the study of different structures including hexagonal lattices with bosonic and fermionic atoms. In particular, recent experiments were able to obtain a complete tomographic image of the Berry curvature of a Bloch band \cite{Duca2015}. No such experiments, that require a fast switching off of the confining (lattice) potentials, are possible in solids.

The question then arises as to what experiments could give direct information on the topological structure of the Bloch wavefunctions in condensed matter systems.
The high resolution angle resolved photoemission  spectroscopy  (ARPES)  has proven to be a powerful tool to measure the dispersion relation of low energy bands,  the band structure of dynamically driven systems (Floquet spectrum) \cite{Wang2013a,Mahmood2016}, quasiparticle lifetimes and even the chiral nature of the electronic states in graphene systems \cite{Liu2011}.  In the latter case, ARPES experiments show that the intensity patterns have an angular dependence that give direct information of the Berry's phase. This is due to the fact that graphene's wavefunctions are spinors corresponding to the pseudo-spin associated with the two sublattices of the hexagonal structure.  Then, close to the Dirac points, the  pseudo-spin  is parallel to the crystal momentum leading to a nontrivial Berry phase of $\pi$.  Similar results are obtained in bilayer graphene where the winding angle is $2\pi$. 
However, neither the band structure nor the Berry phase around the Dirac cones provide enough information to characterize the topological structure of the bands.  

In what follows we show that using a pump and probe setup in graphene and graphene-like systems, photo-electrons can unveil topological phase transitions, {\it{i.e.}} they can unambiguously detect changes in the Chern numbers. On the one hand this is possible due to the structure of the dipole matrix elements linked to the excitation of the electrons at the $\pi$-bands of graphene. On the other hand, although Chern numbers involve the Berry curvature of all $k$-points  in the BZ, the largest contribution comes from  two {\it{hot spots}} --- the Dirac points. As we show below, detailed analysis of the intensity of photo-electrons coming from the corners of the BZ gives the required information to identify topological transitions.
 
Pump and probe experiments consist in coupling the system to an electromagnetic pump pulse followed by a short photo-exciting  ARPES pulse. We consider spatially homogeneous pump pulses of circularly polarized light of frequency $\Omega$. Typical duration  of the pump pulse is $\delta t_\mathrm{pump}\sim250$ fs.
 The photo-excitation due to the probe pulse occurs during the pumping time, being its duration  $1/\Omega<\delta t_\mathrm{probe}\ll\delta t_\mathrm{pump}$  and its polarization either linear or circular.

\begin{figure}[t]
\includegraphics[width=0.98\columnwidth]{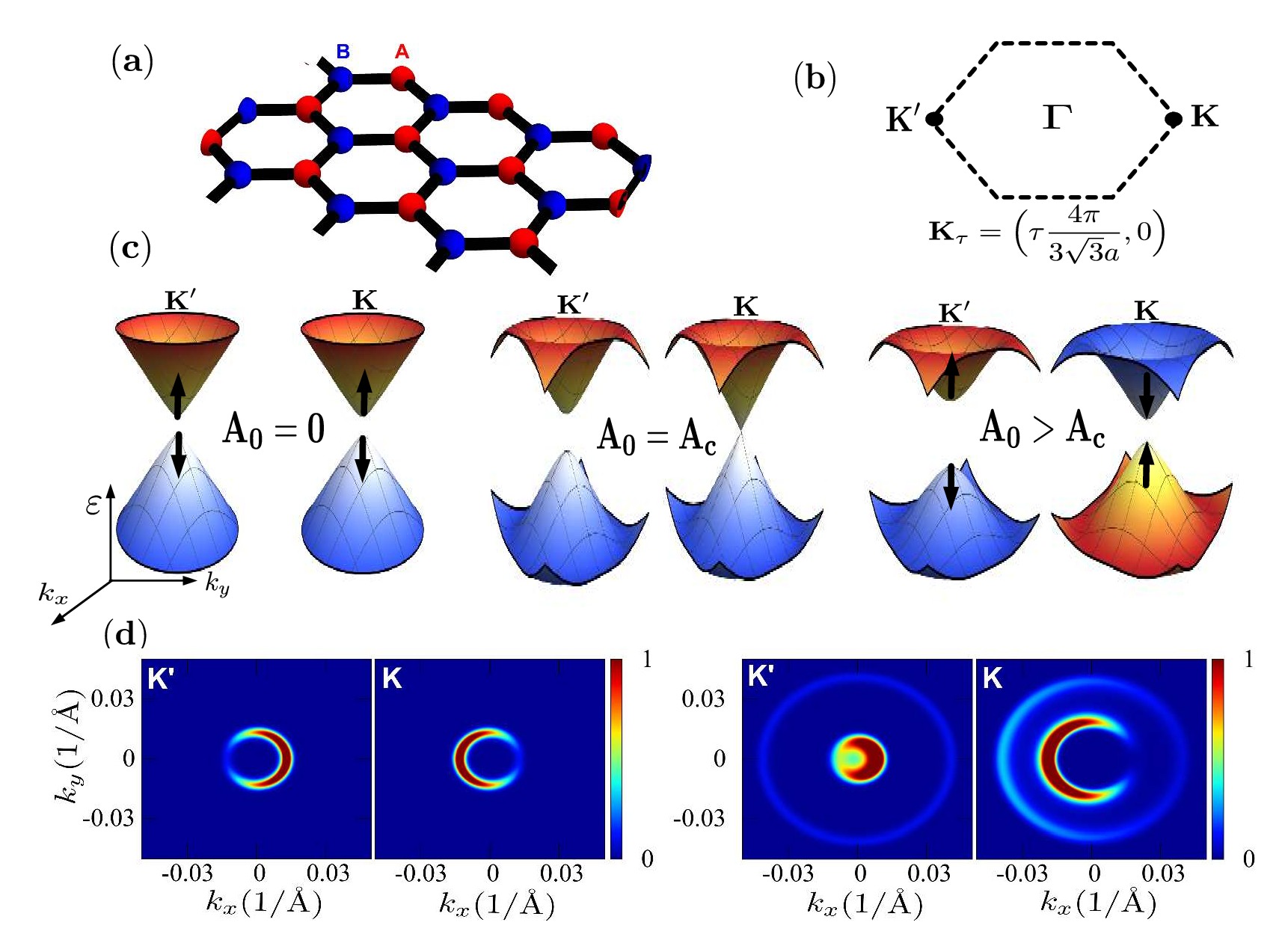}
\caption{The hexagonal lattice with two sites ($A$ and $B$) of the unit cell (a) and the corresponding Brillouin zone (b). In (c) the Floquet band structure near a Dirac point with the amplitude of the vector potential $A_{0}= 0, A_{c}$ and $A_{0}>A_{c}$ are shown; the color of the conduction and valence bands indicate the orientation of the pseudo spin at the Dirac points. (d) The photoemission intensities at constant energy without the pump pulse (left) and with a high energy ($\beta=0$, see text) linearly polarized  probe pulse (right) show a dichroism characteristic of the chiral states; the only effect of the radiation in this configuration is to reveal the change of the constant energy surface at the $K$ and $K^{\prime}$ points and the appearance of Floquet replicas.}
\label{fig1}
\end{figure} 

It is instructive to start our analysis with a simple model of a gaped graphene with a mass term. The  pump pulse is described by a vector potential $\bm{A}(t)$ so that,  for the crystal momentum $\bm{k}$ close to the $K$ or $K^{\prime}$ points of the BZ, the Hamiltonian reads 
\begin{equation}
\mathcal{H}_{\bm{k}\tau}=v_{f}\bm{\Pi}_{\bm{k}\tau} \cdot \bm{\sigma}+\Delta\, \sigma_{z}\,,
\label{eq1}
\end{equation}
where $\bm{\Pi}_{\bm{k}\tau}=(\tau[\hbar k_{x}+e A_{x}(t)],\hbar k_{y}+e A_{y}(t) )$,  $\tau =\pm$ corresponds  to the $K$ or $K^{\prime}$ Dirac points, respectively, and the components of $\bm{\sigma}$ are the Pauli matrices---in our notation the up (down) pseudo-spin  corresponds to the $A$ ($B$) sublattice.
This model describes the band structure of graphene (with $\Delta=0$) and of  silicene or germanene where the mass gap $\Delta$ can be induced by an external electric field \cite{Ezawa2012b} as well as a variety of 2D materials and artificial structures \cite{Wehling2014a}.

Before including the full time dependence of the pump pulse, we consider a circularly polarized monochromatic radiation described by $\bm{A}(t)=\mathrm{Re}(\bm{A}_0\, e^{i \Omega t})$ with $\bm{A}_0=A_{0}\,(\hat{\bm{x}}-i\,\hat{\bm{y}})$.  The time dependent Schr\"odinger equation can be solved in the frame of the Floquet theory \cite{Shirley1965,Sambe1973,Grifoni1998a,Kohler2005}. For states with wavevector close to the Dirac points, to the lowest order in the field amplitude $A_{0}$,  the system is described by the following Floquet Hamiltonian 
\begin{equation}
\mathcal{H}^{F}_{\bm{k}\tau}=\left( \begin{array}{cc} \tilde{\Delta}_{\tau} &
\hbar v_f(\tau k_{x}-i k_{y})\\\hbar v_f(\tau k_{x}+i k_{y}) & -\tilde{\Delta}_{\tau}
\end{array} \right)\,,
\label{eq2}
\end{equation}
with 
\be
\tilde{\Delta}_{\tau}=\Delta-\frac{\tau e^{2}v_f^2 A^{2}_{0}}{\hbar\Omega-\tau\Delta}\,,
\ee
and Floquet quasi-energies $\varepsilon_{k\tau}=\pm \sqrt{{\tilde{\Delta}_{\tau}}^2+(\hbar v_f k)^{2}}$. This solution shows that at the $K$ point ($\tau = +$) the gap decreases as the field amplitude increases, it closes at a critical value $A_{c}=\sqrt{\Delta(\hbar\Omega-\Delta)}/ev_f$ and increases again for $ A_{0}> A_{c} $. On the other hand, the gap at $K^{\prime}$ increases monotonously \cite{Ezawa2013}. Reversing the sense of rotation of the electromagnetic field changes the Dirac point at which the gap closes. 
This phenomena, known as band inversion, is accompanied by a change of the nature of the Floquet wavefunctions at the corners of the BZ. While for $A_{0}<A_{c}$ the wavefunctions of the conduction band for both cones at ${\bm{k}=0}$ are localized on the $A$ sublattice (${\it{i.e.}}$ their pseudo-spin is up), for $A_{0}>A_{c}$ the wavefunction at $K$ lies on sublattice $B$ (down pseudo-spin) as shown in Fig. \ref{fig1}.
 
The band inversion with the closing of the gap at the critical field amplitude signals a topological phase transition \cite{Ezawa2013}. Indeed the Berry curvature of the Floquet states around $K$ and $K^{\prime}$ is given by  
\be
\Omega_{\bm{k}\tau}= -\frac{\tau (\hbar v_f)^2 \tilde{\Delta}_{\tau}}{2 ((\hbar v_{f} k)^{2}+{ \tilde{\Delta}_{\tau}}^{2})^{3/2}}\,,
\ee
and gives a contribution to the Chern number of the valence band ${\mathcal{C}}_{\tau}=-\tau \mathrm{sign}( \tilde{\Delta}_{\tau})/2$~\cite{Xiao2010}. Hence, to this order in the field amplitude,  ${\mathcal{C}}_{+}+{\mathcal{C}}_{-}$ changes from $0$ for $A_{0}<A_{c}$ to $1$ for  $A_{0}>A_{c}$. 
When considering the full tight-binding Hamiltonian of graphene, it can be shown that in undoped nanoribbons the phase having ${\mathcal{C}}_{+}+{\mathcal{C}}_{-}=0$
behaves as a (normal) insulator. In this phase the gap is preserved with non protected helical edge states laying close to the bottom and top of the energy gap. Conversely, in the phase where ${\mathcal{C}}_{+}+{\mathcal{C}}_{-}=1$
the gap is bridged by topologicaly protected chiral edge states and the system becomes a TI \cite{Ezawa2015}.  

We are now in position to address the problem of how the intensity and angular dependence of the ARPES distinguishes the two different topological phases. The photo-excitation process is described by the Hamiltonian  
\be
\mathcal{H}_{w}(t)=w(t)\sum_{\bm{p} \alpha}(M_{\bm{p} \alpha} a^{\dagger}_{\bm{p}}c_{\alpha^{}}+M_{\bm{p} \alpha}^{*} c_{\alpha}^{\dagger}a^{}_{\bm{p}})\,,
\ee
where $w(t)$ describes the time profile of the probe pulse, $a^{\dagger}_{\bm{p}}$ creates a photo-electron with total momentum $\bm{p}$ and $ c_{\alpha}$ annihilates an electron at the sample with quantum numbers $\alpha$.   
Assuming that the probe pulse $w(t)$ acts in the time interval $[t_{0}, t_{1}]$, the total photo-electron distribution obtained after the probe is given by~\cite{Liu2011,Braun2015,Sentef2015}
\begin{widetext}
\begin{equation}
I(\bm{p})=\sum_{m} f(E_{m})\left|\sum_{\alpha}{\int_{t_{0}}^{t_{1}}} M_{\bm{p} \alpha}\, e^{i \varepsilon_{\bm{p}} t^{\prime}/\hbar} w(t^{\prime}) \langle 0|c_{\alpha}\,{\mathcal{U}_{\bm{p}}}(t^{\prime},-\infty)|\psi_{\bm{p}}^m\rangle \,dt^{\prime}\right|^{2}\,. 
\label{eq3}
\end{equation}
\end{widetext}
Here $\varepsilon_{\bm{p}}$ is the energy of the photo-electron, $\ket{\psi_{\bm{p}}^m}$ is a state of the system in equilibrium with energy $E_{m}$, ${\mathcal{U}_{\bm{p}}}(t^{\prime},-\infty)$ is the time evolution operator including the effect of the pump pulse and $f(E)$ is the Fermi-Dirac distribution. The structure of the dipolar matrix elements $M_{\bm{p} \alpha}$ has been discussed in Ref. \cite{Liu2011} for a probe pulse described by a vector potential with a polarization vector given by $\bm{P}_{A}=\cos\chi\, \hat{\bm{x}} - i \sin\chi\, \hat{\bm{y}}$. Using as a complete basis the eigenstates of the unperturbed system ($\alpha=\bm{k},\tau,\pm$ where $\pm$ indicate the valence and conduction bands, respectively) and setting $\bm{p}=\hbar (\bm{K}+\bm{k})$ or $\bm{p}=\hbar (\bm{K}^{\prime}+\bm{k})$ we have $M_{\bm{p} \alpha}\equiv M^{\pm}_{k \tau}=\langle \psi_{f}|\bm{P}_{A} \cdot \bm{p} |\psi^{\pm}_{\bm{k} \tau}\rangle$ with $\psi_{f}$ the wavefunction of the final photo-electron state~\cite{Liu2011}.  These matrix elements are given in terms of the dipole transition matrix elements $\zeta_{x}=\bra{\psi_f}{\bm{p}}_{x}\ket{{\bm{k}}A}=\bra{\psi_f}{\bm{p}}_{x}\ket{{\bm{k}}B}$ and  $\zeta_{y}=\bra{\psi_f}{\bm{p}}_{y}\ket{{\bm{k}}A}=-\bra{\psi_f}{\bm{p}}_{y}\ket{{\bm{k}}B}$ for the $x$ and $y$ components of the probe pulse respectively (see supp. information  \cite{supp}). In the expressions above,  $\ket{{\bm{k}}A}$ and $\ket{{\bm{k}}B}$ are the Bloch wavefunctions of the $A$ and $B$ sublattices, respectively. 

Recent experiments~\cite{Liu2011} showed that for graphene the ratio  $\zeta_{x}/\zeta_{y}=\lambda e^{i \beta}$ depends on the frequency of the photo-emitting probe pulse. For high energies ($\sim 30 $~eV) $\lambda$ is on the order of one and $\beta \simeq 0$ while for lower energies ($\sim 20 $~eV)  $\beta \simeq \pi/2$. In the former case  the momentum distribution of the photo-electrons gives valuable information on the Berry phase and has been analyzed in detail in Ref. \cite{Liu2011} and in subsequent works in the absence of the pump perturbation~\cite{Hwang2014,Hwang2015}.  In this case a simple calculation gives the following photo-electron distribution due to electrons with quantum numbers $\bm{k},\tau,\pm$,
\begin{equation}
I^{\pm}_{\bm{k}\tau} \propto |M^{\pm}_{\bm{k}\tau}|^{2}=1\pm \frac{\tau\hbar v_{f} k}{\sqrt{ {\Delta}^{2}+(\hbar v_{f}k)^{2}}}  \cos(\theta_{\bm{k}}+2\tau \chi) \,,
\label{Ibeta0}
\end{equation}
with 
$\theta_{\bm{k}}=\arctan(k_{y}/k_{x})$. 

\begin{figure}[t]
\includegraphics[width=\columnwidth]{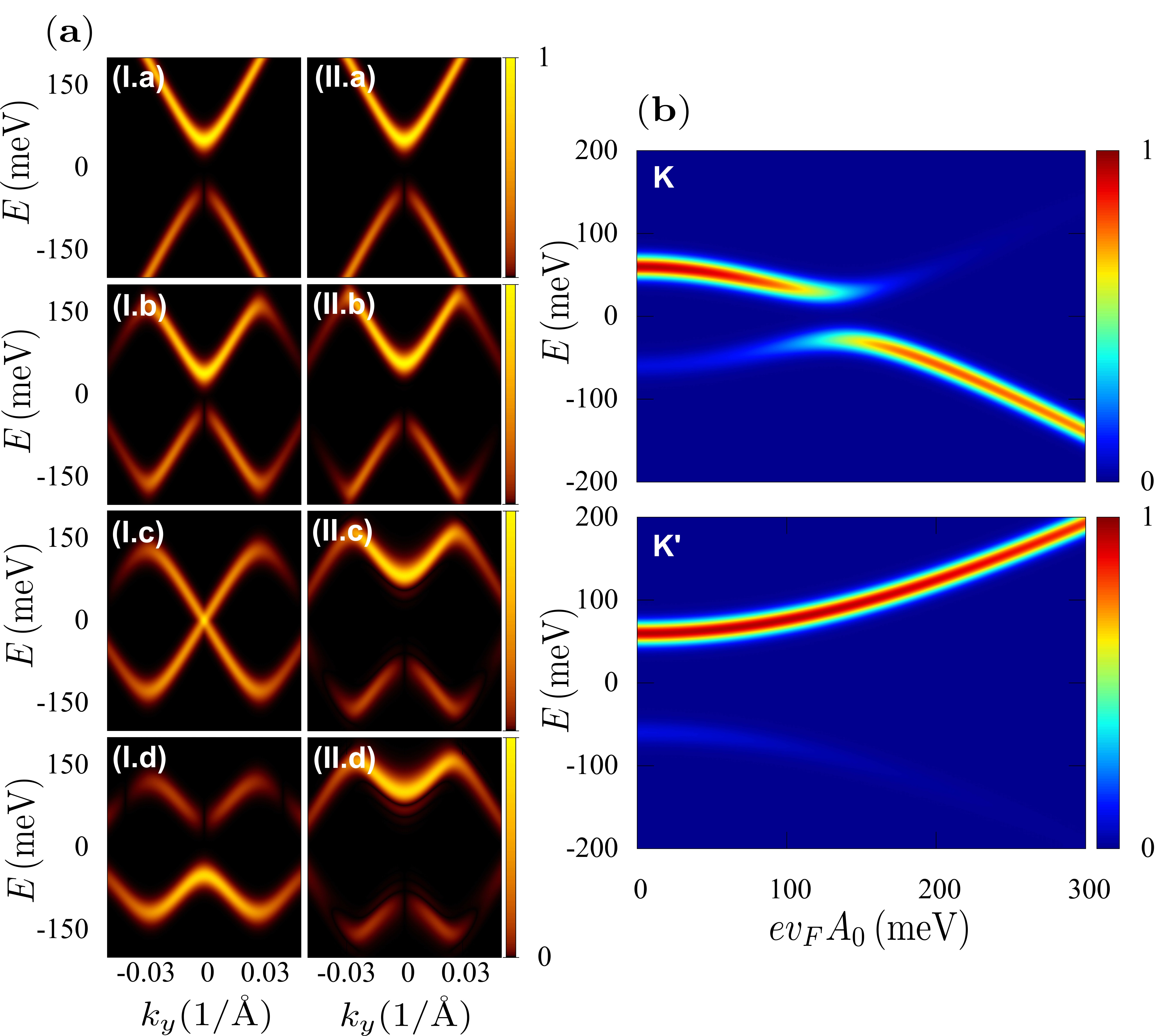}
\caption{(a) ARPES intensity from states close to the $K$ (left column) and $K^{\prime}$ (right column) cones; the radiation intensity increases from top to bottom. These results correspond to a circularly polarized pump and probe pulses with $\beta=\pi/2$, the temporal duration of the former being of $350\,$fs and the latter of $\delta t_{\text{probe}} = 50\,$fs. The chemical potential has been taken at $0.5$~eV to appreciate the intensity changes of photo-electrons from both the valence and the conduction bands. In (b) we show the intensities, from states with wavevector $\bm{k}$ slightly shifted from the Dirac points, as function of the radiation intensity. At the critical value of $A_ {0} = 130\,$meV the the ARPES intensity around $K$ is transferred from the conduction to the valence band.}
\label{fig2}
\end{figure}  
\begin{figure*}[t]
\includegraphics[width=\textwidth]{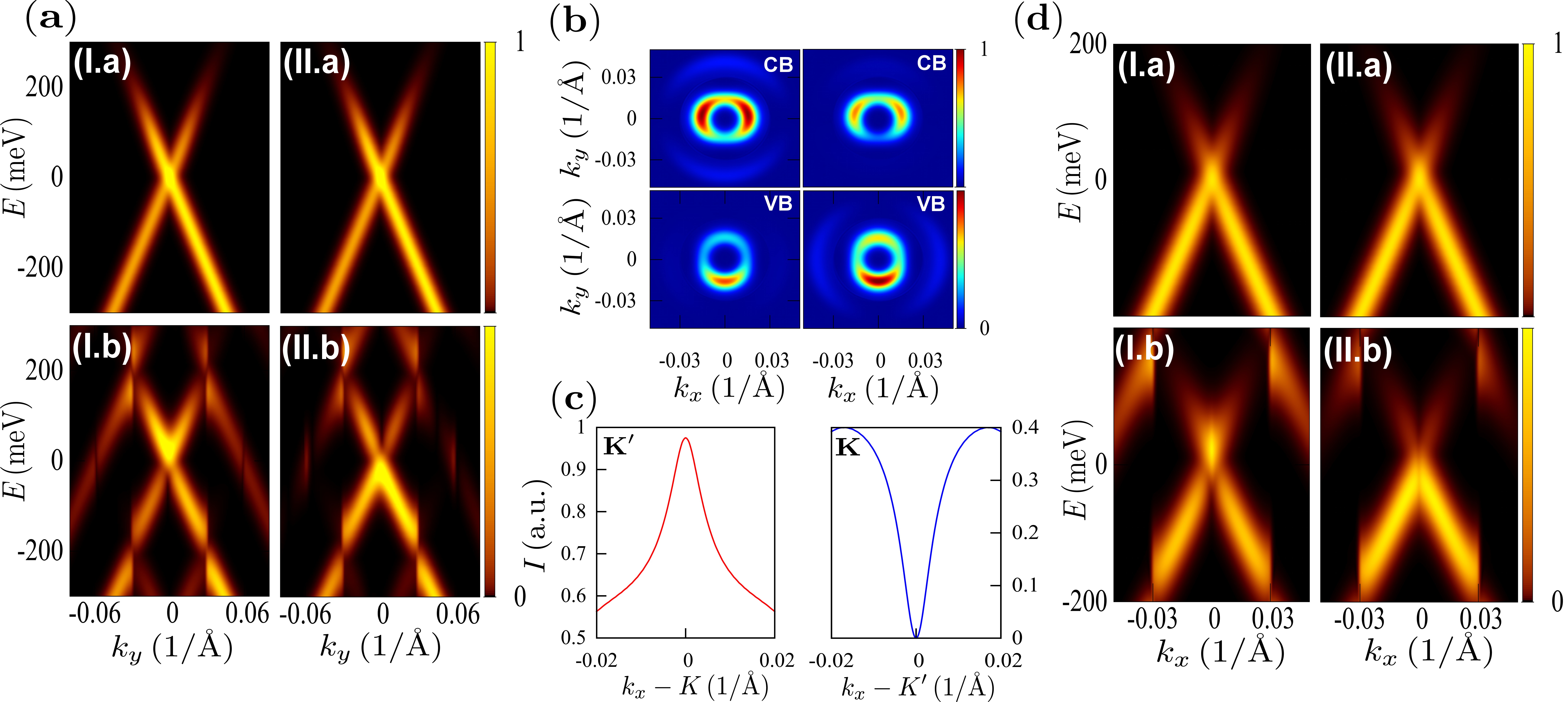}
\caption{ARPES intensity for graphene ($\Delta=0$) and $\beta=0.4\pi$ corresponding to x-ray energies of the probe pulse of $20 $~eV. The pump and probe pulses are circularly polarized, the former with a time domain width of $250\,$fs and the latter with $20\,$fs. (a) Cuts of the ARPES intensity along $k_{y}$ for graphene in equilibrium (upper panels) and irradiated graphene (lower panels) around $K^{\prime}$ (right panels) and $K$ (left panels). (b) Constant energy cuts at $100 $~meV  (CB) and at $-100 $~meV  (VB) around $K^{\prime}$ (right panels) and $K$ (left panels). Note the small dichroism obtained with $\beta=0.4\pi$, contrary to the case of linearly polarized probe, now the dichroism around $K$ and $K^{\prime}$ has the same symmetry. (c) Intensity of the valence band photo-electrons along $k_{x}$ from states close to the  $K^{\prime}$ and $K$ points. The maximum at $K^{\prime}$ and the minimum at $K$ signals the non trivial topology of the Bloch wavefunctions of this band. (d) Cuts of the photoemission spectrum for graphene along $k_x$ with chemical potential fixed at $\mu=0$ eV in equilibrium (upper panels) and irradiated (lower pannels) around $K^{\prime}$ (right panels) and $K$ (left panels).}
\label{fig3}
\end{figure*}  

Photons with different polarization selectively excite electrons in the BZ generating a marked dichroism. This is reflected in the angular dependence of constant energy maps of $I^{\pm}_{\bm{k}\tau}$ close to the $K$ point. The angular dependence of the photo-electron distribution highlights the chiral nature of the initial states and gives direct information of the winding  phase $\theta_{\bm{k}}$. Similar results are obtained with a pump pulse as shown in Fig. \ref{fig1}(d). To lowest order in the pump amplitude, the photo-electron intensities are given by Eq.~(\ref{Ibeta0}) where now $\Delta$ is to be replaced by $\tilde{\Delta}_{\tau}$. Then the pump changes the band structure as suggested by the lowest order Floquet Hamiltonian [cf. Eq.~(\ref{eq2})] and the  closing of the mass gap at $K$ can be observed. However, under these conditions the ARPES spectrum cannot distinguish the two different topological phases. In fact, for $\beta=0$ the photo-electron distribution $I^{\pm}_{\bm{k} \tau}$ is independent of the sign of the mass term, which means that the intensity pattern remains invariant under a change in the orientation of the pseudospin along the $z$ axis. Although the ARPES can detect the closing and reopening of the gap at one of the Dirac points as the amplitude of the pump pulse increases, this cannot be unambiguously assigned to a band inversion. In particular in graphene where $\Delta=0$ the gaps at the two Dirac point are identical and the photo-electron intensities are insensitive to the sign of $\tilde{\Delta}_{\tau}$.  

Interestingly, when $\beta\simeq\pi/2$, a situation experimentally observed for $\hbar \omega\sim 20$~eV, the ARPES spectrum changes at the critical amplitude of the pump pulse allowing for a clear identification of the topology of the Floquet bands.  Before presenting the numerical results we may get some insight into the problem by evaluating the photo-electron distribution
using again the lowest order Floquet Hamiltonian. For $\lambda=1$ and $\beta=\pi/2$ this approach gives
\begin{equation}
I^{\pm}_{\bm{k}\tau} \propto 1\pm [\sin(\tilde{\phi}_{k\tau}) \cos(\theta_{\bm{k}})\cos(2 \chi)+   \cos(\tilde{\phi}_{k\tau}) \sin(2 \chi)]\,,\\
\label{Ibetapi2}
\end{equation}
with $\cos (\tilde{\phi}_{k\tau})=\tilde{\Delta}_{\tau} /\sqrt{ {\tilde{\Delta}_{\tau}}^{2}+(\hbar v_{f}k)^{2}}$ and $\sin (\tilde{\phi}_{k\tau})=\tau \hbar v_{f} k/\sqrt{ {\tilde{\Delta}_{\tau}}^{2}+(\hbar v_{f}k)^{2}}$. This simple result makes apparent that the ARPES spectrum for non-linear polarization of the probe ($\chi \ne 0$ or $\pi /2$) depends on the sign of the mass term. Consequently, the topological transition is manifested as a change in the amplitude of the photo-electron intensities showing different behaviors at the $K$ and $K^{\prime}$ cones. Under this choice of parameters it is possible to generate a photo-electron distribution with purely A or B character, \textit{i.e.} to selectively photo-emit states with different pseudospin polarization along the $z$ axis. Equation (\ref{Ibetapi2}) also shows that the dichroism depends on the helicity of the probe. Defining the dichroism factor $D$ as the normalized maximum angular variation of the photoemission intensity along a constant energy curve we obtain 
\be
D^{\pm}=\frac{\sin(\tilde{\phi}_{k\tau}) \cos(2\chi)}{1\pm \cos(\tilde{\phi}_{k\tau}) \sin(2 \chi)}\,.
\ee
For a circularly polarized probe pulse ($\chi=\pi/4$) we have that $D^{\pm}=0$ and the information on the Berry phase is lost---the constant energy cuts of the photo-electron distribution are angle independent . However, the intensity of the photocurrent coming from the valence and conduction bands clearly shows the topological structure of the wavefunctions. 

This is shown in Fig. \ref{fig2} where the numerical simulation with the full time dependence of the pump and probe pulses are presented. 
The figure was obtained by fixing the chemical potential at a high energy (high doping) in order to show the photo-electron intensities corresponding to the valence and conduction bands in a wide energy range.  The circularly polarized probe pulse acts at the centre of the pump pulse and its width in the time domain was chosen to be $\delta t_\mathrm{probe}=50\,$fs to have a good energy resolution of the Floquet bands. The results clearly show that near the $K$ point, the maximum intensity of the photo-electron distribution changes from the conduction to the valence band at the critical amplitude $A_{c}$. This change is a consequence of the sign change of ${\tilde{\Delta}_{\tau=+}}$ and is linked to a change of ${\mathcal{C}}_{+}+{\mathcal{C}}_{-}$.

To be more specific, we now present results for the case of graphene with realistic parameters.  We used the experimentally observed value of the phase $\beta=0.4\pi$, the chemical potential is set either at $\mu=100 $~meV or $\mu=0 $~meV, the frequency of the pump pulse is $\hbar\Omega=400 $~meV and the probe pulse is circularly polarized. These conditions generate small dichroism although its symmetry is different from that observed with $\beta =0$: note that in cuts along $k_{y}$ and in the absence of the pump pulse the lines with negative velocity in Fig. \ref{fig3}(a) are more intense around both the $K$ and $K^{\prime}$ points. 
The circularly polarized pump pulse with frequency $\Omega$ also opens gaps at the Floquet zone boundary ($\hbar\Omega / 2$) that are detected the by the ARPES spectrum~\cite{Wang2013a}. The second order gap at the zone centre (zero energy) is not clearly observed due to the moderate amplitude of the pump and the width of the ARPES lines, however the intensities of the lines corresponding to the conduction band show a marked different behavior at the two Dirac points as illustrated in Fig. \ref{fig3}(c). This behavior shows that the Berry curvature $\Omega_{k \tau}$ defined above has the same sign for the two cones leading to a non-zero Chern number. This effect is also present when the chemical potential is fixed at $\mu=0\,$eV as shown in Fig. \ref{fig3}(d), where the photoemission spectrum is presented along the $k_x$ direction in order to disregard asymmetries due to the dichroism generated by the probe polarization.

In finite systems, the Floquet zone boundary gaps are bridged by topologically protected edge states. In $k$-space, the edge states are confined arround the $K$ and $K^{\prime}$ points and their existence can be inferred by evaluating the Chern numbers with the Floquet bands~\cite{Rudner2013,Usaj2014a,Perez-Piskunow2015}.  The wavefunctions in the time domain clearly show that for those states bridging the zone-boundary gap the pseudospin oscillates with frequency $\Omega$ with its time average value on the $xy$-plane. The topological structure of these states, described by the above mentioned Chern numbers, is a real dynamical effect~\cite{FoaTorres2014}. As the ARPES probe pulse averages on a time scale of the order of $\delta t_\mathrm{probe}\gg1/\Omega$,  the photo-electrons can hardly carry some information on the topological nature of states at the zone-boundary gap. 

It is worth mentioning that, as recently shown in Ref. [\onlinecite{DAlessio2015}], the Chern number of a pure state (Slater determinant) cannot be changed by a unitary transformation, that is, the Chern of an initial state remains unaltered during the pump pulse. This fact of course does not prevent modifications of the band structure, the Floquet spectrum, and in particular the presence of the band inversion phenomena. In ARPES experiments with the appropriate  energy and polarization, the interference of the dipole transitions allows for a clear identification of the different topological phases as revealed by the band inversion effect. With the help of a band structure model, that for graphene is well established, the ARPES intensity profiles allows to determine amplitude and phases of the wavefunctions for states close to the $K$ and $K^{\prime}$ points of the BZ and to reconstruct the Berry curvature around these hot spots.   

The case of bilayer graphene, with a rather different band structure, is also interesting. The system has four $\pi$-bands, two of them, with parabolic dispersions, touch each other at the Dirac points and a gap can be opened and controlled by a perpendicular electric field. The other two bands lie at about $0.3$~eV from the Dirac points. In the presence of the pump pulse these extra bands generate Floquet replicas that partially cover up the low energy ARPES spectrum making it much more intricate. Nevertheless, as the pump amplitude increases the topological transition evidenced by the band inversion phenomena can be clearly observed (see Supplementary Information~\cite{supp})    

In summary, we have shown that ARPES can give clear information on the topology of Floquet bands of graphene and graphene-like structures. This information is given by the intensity of the ARPES profiles of the bands close to the $K$ and $K^{\prime}$ points of the BZ. While in the topological trivial phase the intensities due to photo-electrons from the valence or conduction bands are similar at the two Dirac points, in the non-trivial phase the intensities of the valence and conduction bands are different and opposite at $K$ and $K^{\prime}$. This change signals a modification of the Berry curvature around these points with a consequent variation of the Chern numbers. To observe the effect the dipole transition matrix elements $\zeta_{x}$ and $\zeta_{y}$ should have a different phase $\beta$. It has been experimentally shown that in graphene $\beta$ can be controlled with the photon energy of the probe pulse.  
This observation opens the road for a spectroscopic study of the topological properties of the \textit{bulk} wavefunctions of these 2D materials.

We acknowledge financial support from PICTs 2013-1045 and Bicentenario 2010-1060 from ANPCyT, PIP 11220110100832 from CONICET and grant 06/C415 from SeCyT-UNC. GU acknowledges support from the ICTP associateship program and  thanks the Simons Foundation.
\bibliographystyle{apsrev4-1_title}
\bibliography{paper,note}
\newpage
\begin{widetext}

\section{Supplementary information for ``Photoelectrons unveil topological transitions in graphene-like systems" }
\section{The time evolution operator}
The total Hamiltonian is written as $\mathcal{H}(t)={\sum_{\mathbf{k}\tau}}\mathcal{H}_{\mathbf{k}\tau}(t)$ with 
\begin{equation}
\mathcal{H}_{\mathbf{k}\tau} (t)= v_f {\Pi}^{x}_{\bm{k}\tau}(t)\sigma_x + v_f {\Pi}^{y}_{\bm{k}\tau}(t)\sigma_y + \Delta\sigma_z = \mathbf{d}_{\mathbf{k}\tau}(t)\cdot{\boldsymbol{\sigma}}
\label{eq1}
\end{equation}
here $\mathbf{d}_{\mathbf{k}\tau}(t) = \Big(\tau v_f(\hbar k_x + e A_x(t)),\hbar v_f k_y + e v_f A_y(t),\Delta\Big)$, where the pump vector potential $\bm{A}(t) = \Re[\bm{A_0}(t)e^{i\Omega t}]$ has been introduced via minimal coupling ($\Pi_{\bm{k}\tau}^{\nu} = \hbar k_{\nu} + eA_{\nu}(t)$) with an envelope function $\bm{A_0}(t)$ and $\bm{\sigma} = (\sigma_x, \sigma_y, \sigma_z)$.

The time evolution operator from an initial time $t_{i}$  to time $t$ acting on a state with quantum numbers $\mathbf{k}$ and $\tau$ is 
\begin{equation}
\mathcal{U}_{\mathbf{k}\tau}(t,t_i) = \mathcal{T}\Big[e^{-\frac{i}{\hbar}\int_{t_i}^{t}\bm{d}_{\mathbf{k}\tau}(t')\cdot \boldsymbol{\sigma}dt'}\Big],
\end{equation}
where $\mathcal{T}$ is the time ordering operator. Using small time intervals $\delta t $ the above integral is approximated as a sum 
\begin{equation}
{\mathcal{U}}_{\mathbf{k}\tau} (t,t_i) = \mathcal{T}\Big[e^{-\frac{i}{\hbar}\sum_{n=1}^{N} \bm{d}_{\mathbf{k}\tau}(t_n)\cdot \boldsymbol{\sigma}\delta t}\Big] \approx \mathcal{T}\Big[\prod_{n} e^{-\frac{i}{\hbar}\bm{d}_{\mathbf{k}\tau}(t_n)\cdot \boldsymbol{\sigma}\delta t}\Big],
\end{equation}
with $t_n = t_i + \frac{2n-1}{2}\delta t$. The last term in the above equation is obtained asuming that $[\mathcal{H}_{\mathbf{k}}(t_n), \mathcal{H}_{\mathbf{k}}(t_n + \delta t)] \approx 0$ for small enough $\delta t$.
Using $[\boldsymbol{\sigma}\cdot\hat{\bm{d}}_{\bm{k}\tau}(t_n)]^{2n} = \mathbf{1}$ and $[\boldsymbol{\sigma}\cdot\hat{\bm{d}}_{\bm{k}\tau}(t_n)]^{2n+1} = \boldsymbol{\sigma}\cdot\hat{\bm{d}}_{\bm{k}\tau}(t_n)$, with $\hat{\bm{d}}_{\bm{k}\tau} = \bm{d}_{\bm{k}\tau}/|\bm{d}_{\bm{k}\tau}|$, the time evolution operator can be written as
\begin{equation}
\mathcal{U}_{\mathbf{k}\tau}(t,t_i) = \mathcal{T}\Big[\prod_{n}\Big\{\cos\Big(|\mathbf{d}_{\mathbf{k}\tau}(t_n)|\frac{\delta t}{\hbar}\Big)\mathbf{1} - i \sin\Big(|\mathbf{d}_{\mathbf{k}\tau}(t_n)|\frac{\delta t}{\hbar}\Big)\boldsymbol{\sigma}\cdot\hat{\bm{d}}_{\mathbf{k}\tau}(t_n)\Big\}\Big].
\end{equation}

To illustrate the effect of the pump pulse on an unperturbed graphene wavefunction $|\Phi_{\bm{k}\tau}^{\gamma}\rangle$ where $\gamma=\pm$ stands for a state in the valence and conduction band respectively, we calculate the probability $P_{\bm{k}}(t)=|\langle\Phi_{\bm{k}\tau}^{\gamma^{\prime}}|\mathcal{U}_{\mathbf{k}\tau}(t,t_i) |\Phi_{\bm{k}\tau}^{\gamma}\rangle|^2$ of finding a final state $|\Phi_{\bm{k}\tau}^{\gamma^{\prime}}\rangle$ at time $t$ with an initial time $t_i$ preceding the pumping.

According to the Floquet theorem when the system is perturbed by circularly polarized radiation of frequency $\Omega$, the Floquet spectrum shows gaps at the Floquet zone centre, with zero energy, and at the Floquet zone-boundary of energy $\hbar \Omega/2$. These energies correspond to wavectors $k=0$ and $k=k_0=\Omega/2 v_f$. Fig. \ref{figS1} shows that even for very short pump pulses, the evolution of the wave functions with $k=0$ and $k=k_0$ considerably differs from those with other values of $k$ away from any anticrossing of the spectrum. In fact for $k=0, k_0$ the system is in a resonant condition, with the pseudospin oscillating between the ${\it{up}}$ and ${\it{down}}$ states with a dominant frequency $\omega$ given by the corresponding Floquet gap. For other values of $k$, an out of resonance condition,  the amplitude of the oscillations decreases and its main frequency is the frequency $\Omega$ of the pump.
\begin{figure}[t]
\includegraphics[width=0.9\columnwidth]{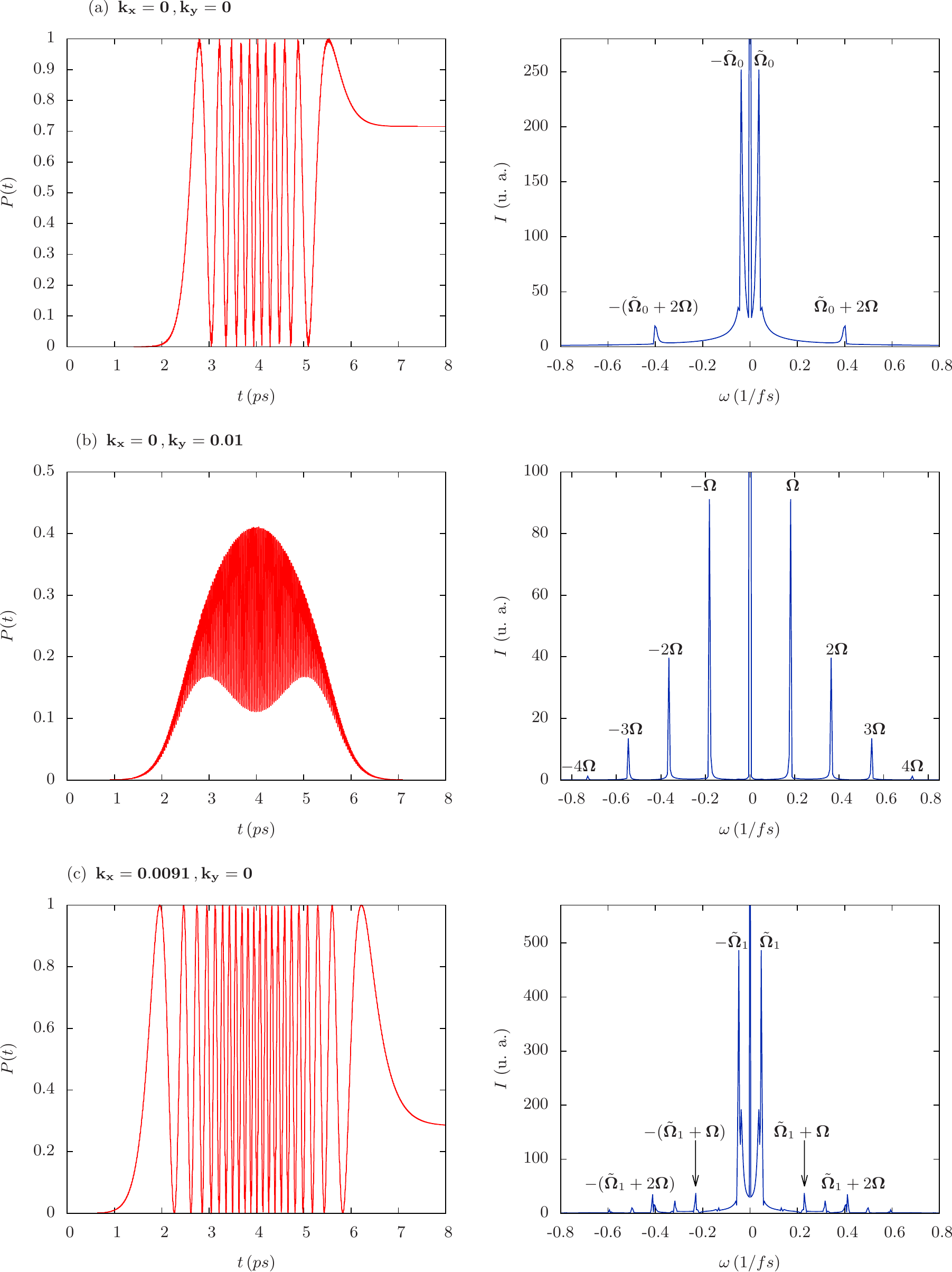}
\caption{Probability $P_{\bm{k}}(t)=|\langle\Phi_{\bm{k}\tau}^{+}|\mathcal{U}_{\mathbf{k}\tau}(t,t_i) |\Phi_{\bm{k}\tau}^{-}\rangle|^2$ of finding a final state $|\Phi_{\bm{k}\tau}^{+}\rangle$ at the conduction band at time $t$ with an initial state $|\Phi_{\bm{k}\tau}^{-}\rangle$ at the valence band at time $t_i$ preceding the pumping for crystal momentum (left pannels) (a) at the Dirac cone K, (b) away from any anticrossing and (c) at $k_0 = \frac{\Omega}{2v_f}$ with their corresponding fourier transform (right pannels).}
\label{figS1}
\end{figure} 

\section{Dipolar Matrix Elements}
The eigenfunctions of the Hamiltonian given by Eq. (\ref{eq1}) with $\Delta\ne 0$ are given by
\begin{equation}
\begin{split}
|{\Psi_{\bm{k}\tau}^{+}}\rangle = \cos\Big(\frac{\phi_{\bm{k}\tau}}{2}\Big)|{\bm{k}},A\rangle + \sin\Big(\frac{\phi_{\bm{k}\tau}}{2}\Big)e^{i\tau\theta_{\bm{k}}}|{\bm{k}},B\rangle\\
|{\Psi_{\bm{k}\tau}^{-}}\rangle= \sin\Big(\frac{\phi_{\bm{k}\tau}}{2}\Big)|{\bm{k}},A\rangle - \cos\Big(\frac{\phi_{\bm{k}\tau}}{2}\Big)e^{i\tau\theta_{\bm{k}}}|{\bm{k}},B\rangle, 
\end{split}
\label{eigenmass}
\end{equation}
where $|{\bm{k}},A\rangle$ and $|{\bm{k}},B\rangle$ are the Bloch wavefunctions of the $A$ and $B$ sublattice respectively and the $\pm$ index refers to the conduction and valence bands, $\theta_{\bm{k}}$ is the angle formed by $\bm{k}$ and the x-axis, $\cos\Big(\frac{\phi_{\bm{k}\tau}}{2}\Big) = {\tau\hbar v_f |\bm{k}|}/{\sqrt{(\hbar v_f k)^2 + (\varepsilon_{+} - \Delta)^2}}$ and $\sin\Big(\frac{\phi_{\bm{k}\tau}}{2}\Big) = {(\varepsilon_{+} - \Delta)}/{\sqrt{(\hbar v_f k)^2 + (\varepsilon_{+} - \Delta)^2}}$ with $\varepsilon_{+}=\sqrt{\Delta^2 + (\hbar v_f k)^2}$.

The dipolar matrix elements $\langle f|\bm{P}_A\cdot \bm{p}|{\Psi_{\bm{k}\tau}^{\pm}}\rangle$ are given in terms of $\zeta_x = \bra{f}p_x\ket{\bm{k}A} = \bra{f}p_x\ket{\bm{k}B}$ and $\zeta_y =\bra{f}p_y\ket{\bm{k}A} = -\bra{f}p_y\ket{\bm{k}B}$, where the relative signs are due to the symmetries of the graphene lattice. The vector potential describing the probe (ARPES) pulse  is $\bm{A}_{pr}(t) = A_{pr}(t)\Re[e^{i\omega t}\bm{P}_A]$ where $\bm{P}_A = \cos(\chi)\hat{\bm{x}} - i \sin(\chi)\hat{\bm{y}}$ and defining ${\zeta_y}/{\zeta_x} = \lambda e^{i\beta}$ the matrix elements  are
\begin{equation}
\begin{split}
M_{\bm{k}\tau}^{+} \propto 
\cos(\chi)\Big\{\cos\Big(\frac{\phi_{\bm{k}\tau}}{2}\Big) + \sin\Big(\frac{\phi_{\bm{k}\tau}}{2}\Big)e^{i\tau\theta_{\bm{k}}}\Big\} - i\sin(\chi)\lambda e^{i\beta}\Big\{\cos\Big(\frac{\phi_{\bm{k}\tau}}{2}\Big) - \sin\Big(\frac{\phi_{\bm{k}\tau}}{2}\Big)e^{i\tau\theta_{\bm{k}}}\Big\}\\
M_{\bm{k}\tau}^{-} \propto \cos(\chi)\Big\{\sin\Big(\frac{\phi_{\bm{k}\tau}}{2}\Big) - \cos\Big(\frac{\phi_{\bm{k}\tau}}{2}\Big)e^{i\tau\theta_{\bm{k}}}\Big\} - i\sin(\chi)\lambda e^{i\beta}\Big\{\sin\Big(\frac{\phi_{\bm{k}\tau}}{2}\Big) + \cos\Big(\frac{\phi_{\bm{k}\tau}}{2}\Big)e^{i\tau\theta_{\bm{k}}}\Big\},
\end{split}
\end{equation}

The ratio ${\zeta_y}/{\zeta_x} = \lambda e^{i\beta}$  depends on the x-ray energies of the ARPES excitation, experimental values for $\lambda$ and $\beta$ are given in Ref. \cite{Liu2011}. It is important to note that for $\beta = \frac{\pi}{2}$ and $\lambda = 1$, admitting the possibility of photoemitting electrons with a probe pulse with circular polarization ($\chi = \pm \frac{\pi}{4}$), a selective projection of the pseudospin along the $z$ axis is achievable. This means that it is plausible to generate a photoelectron current with entirely A or B character, depending on whether the probe polarization is right or left, respectively. These matrix elements are used for the numerical calculation of the  photoelectron intensity. The numerical results with the full time dependence of the driving pump near the Dirac cones can be interpreted in terms of the approximate expression
\begin{equation}
\begin{split}
I_{\bm{k}}^{\pm}\propto |M_{\bm{k}\tau}^{\pm}|^2 = \cos^2(\chi) + \lambda^2 \sin^2(\chi)\pm \Big\{\sin(\widetilde{\phi}_{\bm{k}\tau})\cos(\theta_{\bm{k}})[\cos^2(\chi) - \lambda^2 \sin^2(\chi)]\\
+ \lambda \sin(2\chi)[\sin(\beta)\cos(\widetilde{\phi}_{\bm{k}\tau}) - \cos(\beta)\sin(\widetilde{\phi}_{\bm{k}\tau})\sin(\tau\theta_{\bm{k}})]\Big\},
\end{split}
\end{equation}
here $\cos(\widetilde{\phi}_{\bm{k}\tau}) = {\widetilde{\Delta}}/{\sqrt{\widetilde{\Delta}^2 + (\hbar v_f k)^2}}$ and $\sin(\widetilde{\phi}_{\bm{k}\tau}) = {\tau\hbar v_f |\bm{k}|}/{\sqrt{\widetilde{\Delta}^2 + (\hbar v_f k)^2}}$. The mass term $\widetilde{\Delta}$ is renormalized by the presence of the circular electromagnetic driving.

The dichroism factor $D_{\pm}(\bm{k})$ is defined as $D_{\pm}(\bm{k})=\{\text{Max}[I_{\bm{k}}^{\pm}(\theta_{\bm{k}})]-\text{Min}[I_{\bm{k}}^{\pm}(\theta_{\bm{k}})]\}/\{\text{Max}[I_{\bm{k}}^{\pm}(\theta_{\bm{k}})]+\text{Min}[I_{\bm{k}}^{\pm}(\theta_{\bm{k}})]\}$ . For $\beta=\pi /2$ this gives
\begin{equation}
D^{\pm}(\bm{k})=\sin(\phi_{\bm{k}\tau}) \cos(2\chi)/(1\pm \cos(\phi_{\bm{k}\tau}) \sin(2 \chi))
\end{equation}
\section{ARPES in Bilayer Graphene}

In the Bernal structure the unit cell of the Bilayer Graphene has four C atoms. Consequently there are four $\pi$-bands, two of them with opposite parabolic dispersions touch each other at the Dirac points. The other two bands are shifted by $\approx 0.3 eV$. These four low energy bands are described by the Hamiltonian  $H=H_{1}+H_{2}+H_{12}$ where the first terms describes the electronic structure of two isolated graphene sheets and the last term the interplane coupling  
\begin{equation}
\begin{aligned}
H_{i}=V(-1)^{i-1}\sum_{\bf{k},\sigma}[a^{\dagger}_{i,\bf{k},\sigma}a_{i,\bf{k},\sigma} + b^{\dagger}_{i,\bf{k},\sigma}b_{i,\bf{k},\sigma}\\
-t(\phi({\bf{k}})a^{\dagger}_{i,\bf{k},\sigma}b_{i,\bf{k},\sigma}+\phi^{*}({\bf{k}})b^{\dagger}_{i,\bf{k},\sigma}a_{i,\bf{k},\sigma})]\\
\end{aligned}
\end{equation}
with $i=1, 2$, $a_{i,\bf{k},\sigma}$ and $b_{i,\bf{k},\sigma}$ destroy electrons with wavector $\bf{k}$ and spin $\sigma$ in sublattices $A$ and $B$ of the $i^{th}$ plane respectively and we have included an electric field perpendicular to the BLG plane described by $V$. The matrix element $t$ corresponds to the intraplane hopping and
\begin{equation}
\phi(\mathbf{k})=e^{iak_{y}}\left[1+2e^{-i\frac{3a}{2}k_{y}}\cos\Big(\frac{a\sqrt{3}}{2}k_{x}\Big)\right]
\end{equation}
with $a=1.42\,\text{\AA}$ the carbon-carbon distance. 

The interplane coupling is described by:
\begin{equation}
H_{12}=\sum_{\bf{k},\sigma}t_{\perp}(a^{\dagger}_{1\bf{k},\sigma}b_{2,\bf{k},\sigma}+b^{\dagger}_{2,\bf{k},\sigma}a_{1,\bf{k},\sigma})
\end{equation}

For each value of the wave-number $\bf{k}$ we have a $4X4$ Hamiltonian $H_{\bf{k}}$ given by
\begin{equation}
H_{\bf{k}}=\left| \begin{array}{cccc}V & -t\phi({\bf{k}}) & 0 & t_{\perp
} \\  -t\phi^{*}({\bf{k}}) &V& 0 &  0\\
  0 & 0&-V &  -t\phi({\bf{k}}) \\  t_{\perp} & 0 &  -t\phi^{*}({\bf{k}}) &-V
 \end{array} \right|
\end{equation}
with wavector $[u^{m}_{A1}({\bf{k}}),u^{m}_{B1}({\bf{k}}),u^{m}_{A2}({\bf{k}}),u^{m}_{B2}({\bf{k}})]^{T}$ and eigenvalues $\varepsilon_{m}({\bf{k}})$. Linearizing around the Dirac cones $K_{\tau}$ and coupling the crystal momentum with radiation via Peierls substitution one can readily obtain a time dependent $H_{\bm{k}\tau}(t)$. The complete evolution of the wavefunction during the driving pulse is obtained numerically by means of the evolution operator. In this case, this time dependent propagator is computed by an exact diagonalization of the hamiltonian at each instant of time:

\begin{equation}
\mathcal{U}_{\bm{k}\tau} = \mathcal{T}\Big[\prod_n \sum_m e^{-i\varepsilon_{\bm{k}m}(t_n)\frac{\delta t}{\hbar}}\ket{\Psi^{m}_{\bm{k}\tau}(t_n)}\bra{\Psi^{m}_{\bm{k}\tau}(t_n)}\Big] = \mathcal{T}\Big[\prod_n \mathbb{P}_{\bm{k}\tau}(t_n)\Big],
\end{equation}
where $\mathbb{P}_{\bm{k}\tau}(t_n) = \sum_m e^{-i\varepsilon_{\bm{k}m}(t_n)\frac{\delta t}{\hbar}}\ket{\Psi^{m}_{\bm{k}\tau}(t_n)}\bra{\Psi^{m}_{\bm{k}\tau}(t_n)}$ and $m$ is the band index. The dipolar matrix elements at each valley take the form
\begin{equation}
M_{\bm{k}m}^{\tau} \propto  \cos(\chi)[u^{m}_{A1}({\bf{k}}) + u^{m}_{B1}({\bf{k}}) + u^{m}_{A2}({\bf{k}}) + u^{m}_{B2}({\bf{k}})] - i\sin(\chi)\lambda e^{i\beta}[u^{m}_{A1}({\bf{k}}) - u^{m}_{B1}({\bf{k}}) + u^{m}_{A2}({\bf{k}}) - u^{m}_{B2}({\bf{k}})].
\label{matbilayer4}
\end{equation}

Taking the limit of $\beta = \frac{\pi}{2}$, $\lambda = 1$ and circular polarization $\chi = \frac{\pi}{4}$ the generated photocurrent has only $\mathcal{A}_1$ and $\mathcal{A}_2$ character. By changing the quirality of the probe polarization to $\chi = \frac{7\pi}{4}$ the radiation field couples with $\mathcal{B}_1$ and $\mathcal{B}_2$ sublattices.

The low energy excitations with crystal momentum arround the $K$ and $K^{\prime}$ points of the BZ can be described by an effective two band Hamitonian, obtained by eliminating the bands that are shifted from the Fermi energy by $t_{\perp}$ by means of a canonical transformation. In the base of the $A2$ and $B1$ orbitals, the effective Hamiltonian for a given wavevector $\bf{k}$ takes the form:

\begin{equation}
\label{Hblg}
H_{\bm{k}\tau}=\bm{h}_{\tau}(\bm{k}) \cdot \bm{\sigma}
\end{equation}
where $\bm{\sigma}$ are the Pauli matrices and

\begin{equation}
\begin{aligned}
h_{x}&=\alpha\left(k_{x}^{2}-k_{y}^{2}\right),\\
h_{y}&=2\alpha \tau k_{x}k_{y},\\
h_{z}&=-V,\\
\end{aligned}
\end{equation}
with $\alpha = \frac{9}{4}\frac{(ta)^2}{t_{\perp}} = \frac{(\hbar v_f)^2}{t_{\perp}}$. This two band effective problem is similar to the graphene with mass model, with the advantage of having the possibility to regulate at will the electric field in order to change the parameter $V$. The quadratic (instead of linear) dispersion of these low energy bands is responsable for a coupling of higher order with Floquet replicas, with a lowest order modification of the bias given by
\begin{equation}
\tilde{V}_{\tau} = V + \tau\frac{(e v_f A_0)^4}{t_{\perp}^2(2\hbar\Omega + \tau V)}.
\end{equation}

The effect of a renormalization of $V$ when the system is irradiated generates a closing gap in the cuasi-energy spectrum at one Dirac point and a corresponding opening at the other, making it plausible to detect the band inversion phenomena, as shown in Fig. \ref{figS2}. The frequency of the pump pulse was set at $\hbar\Omega = 0.5\,$eV in order to neglect the influence of replicas from the high energy bands near the Dirac cones. The ARPES instensity of the effective bilayer two band model is also shown in Fig. \ref{figS3} for each valley. In this case the incident photon radiation was set at $\hbar\Omega = 0.2\,$eV.
\begin{figure}[tb]
\includegraphics[width=0.8\columnwidth]{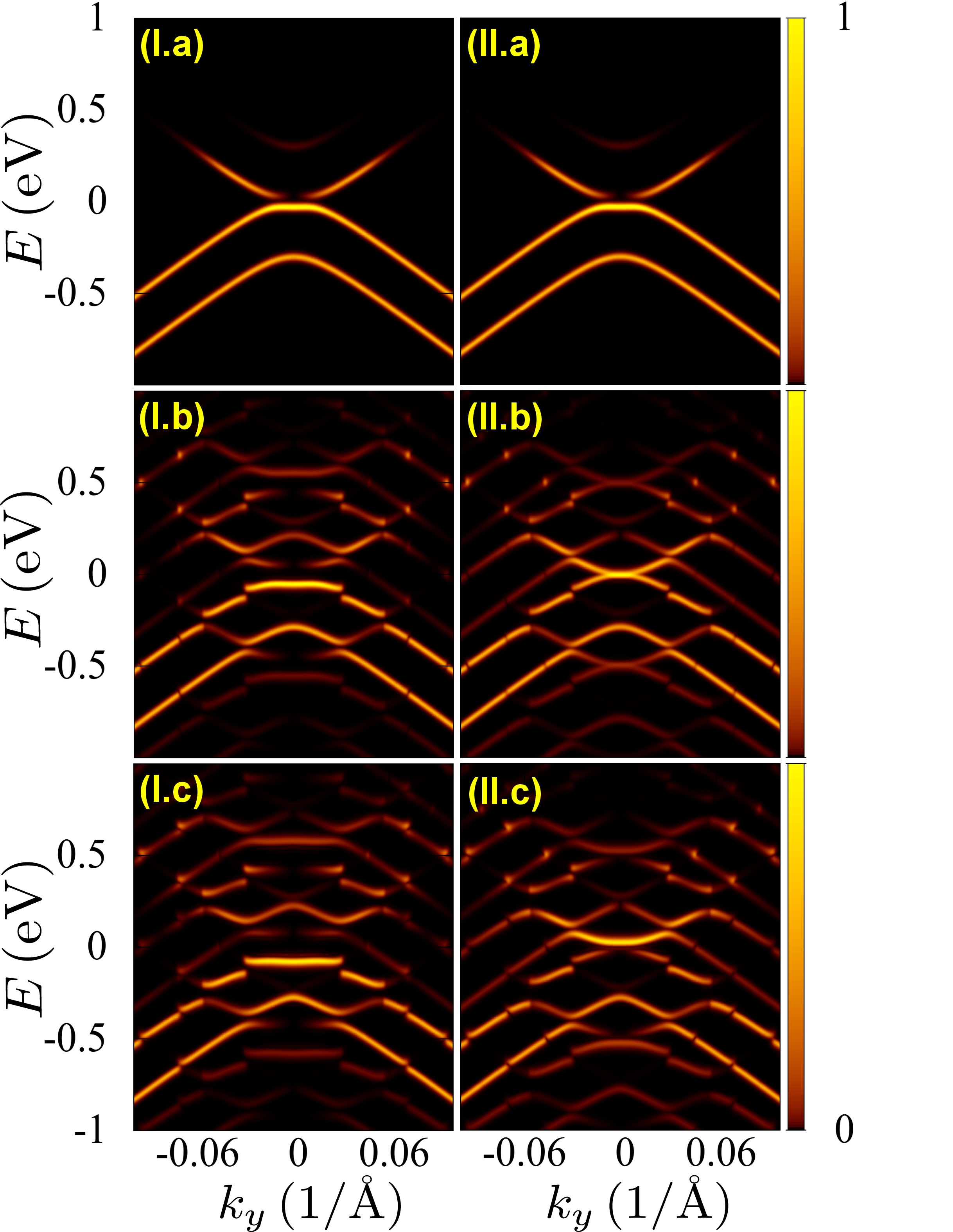}
\caption{ARPES intensity of bilayer graphene from states close to $K$ (left column) and $K^{\prime}$ (right column) Dirac cones; the radiation intensity increases from top to bottom. These results correspond to a circularly polarized pump and probe pulses with $\beta=\pi/2$. The chemical potential has been taken at $0.3$~eV.}
\label{figS2}
\end{figure} 
\begin{figure}[tb]
\includegraphics[width=0.8\columnwidth]{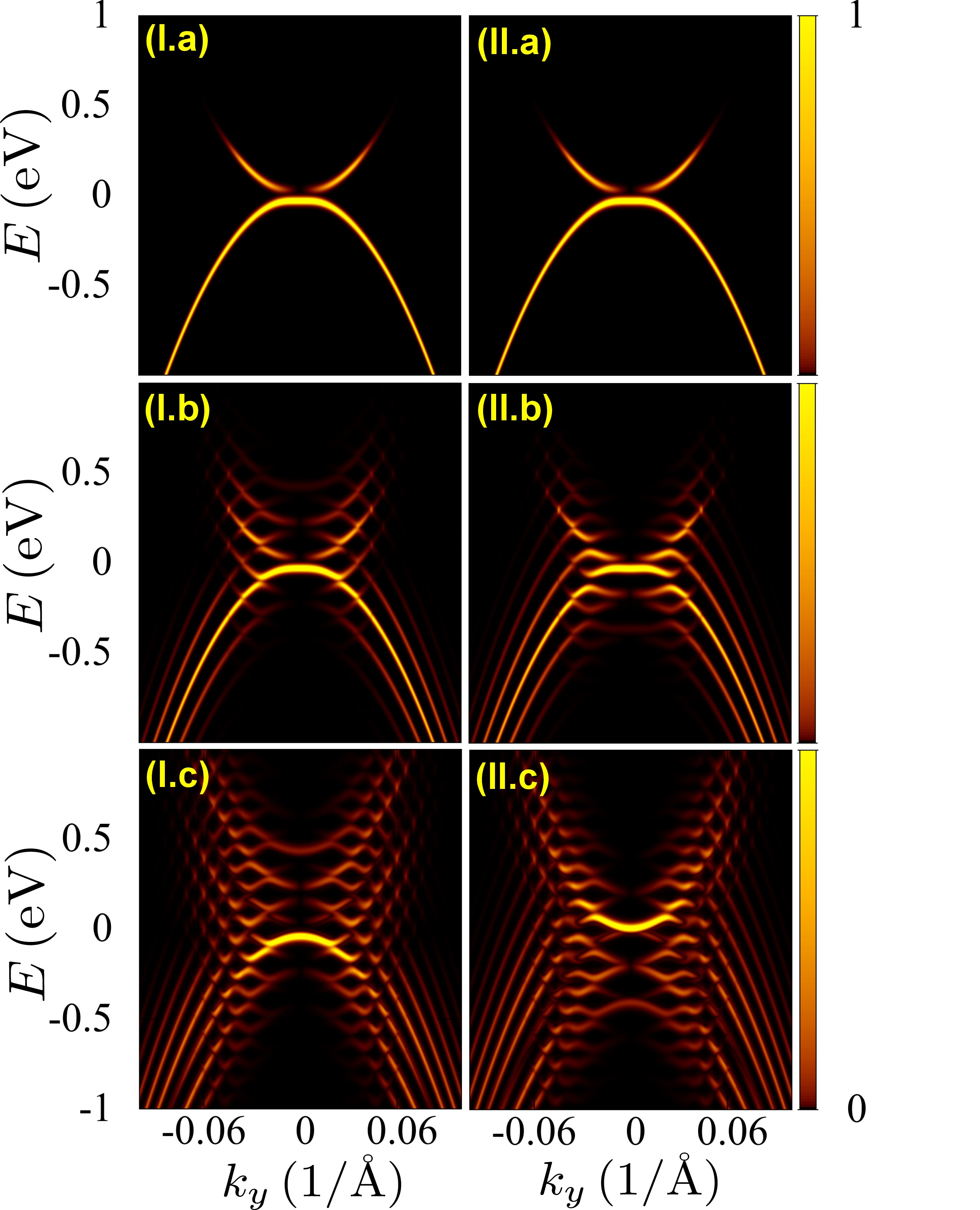}
\caption{ARPES intensity of the effective bilayer two band model from states close to $K$ (left column) and $K^{\prime}$ (right column) Dirac cones; the radiation intensity increases from top to bottom. These results correspond to a circularly polarized pump and probe pulses with $\beta=\pi/2$. The chemical potential has been taken at $0.3$~eV.}
\label{figS3}
\end{figure} 
\end{widetext}

\end{document}